\documentclass[a4paper,11pt,twocolumn]{article}
\usepackage[a4paper,margin=2cm]{geometry} 
\usepackage{lmodern}
\usepackage{booktabs}
\usepackage[table]{xcolor}
\usepackage{graphicx}
\usepackage{latexsym}
\usepackage{amssymb}
\usepackage{float}
\usepackage{balance}
\usepackage{fancyhdr}
\usepackage{textcomp}
\usepackage[version=3]{mhchem}
\usepackage[font={sf,small},labelfont=bf]{caption}
\DeclareCaptionLabelSeparator{vline}{ \boldmath$\vert$ }
\newcommand{\degr}{\ensuremath{\,^\circ}}

\newcommand{\beginsupplement}{
	\setcounter{table}{0}
	\renewcommand{\thetable}{S\arabic{table}}
	\setcounter{figure}{0}
	\renewcommand{\thefigure}{S\arabic{figure}}
}

\usepackage{csquotes}
\usepackage[backend=biber,style=nature,natbib=true,autocite=superscript]{biblatex} 
\AtEveryBibitem{\clearfield{issn}}
\AtEveryCitekey{\clearfield{issn}}
\usepackage[linkcolor = blue!45!black, citecolor = blue!65!black, colorlinks = true, menucolor = blue!65!black]{hyperref}

\usepackage{soul,xcolor} 

\newcommand*{\ins}[1]{{#1}}

\newcommand*{\rem}[1]{}

\title{\flushleft\textsf{Porous translucent electrodes enhance current generation from photosynthetic biofilms}} 

\author{\textsf{Tobias Wenzel\,$^1$, Daniel Haertter\,$^{1,2}$, Paolo Bombelli\,$^3$, Christopher J. Howe\,$^3$*, Ullrich Steiner\,$^{4}$*}}
\date{}

\begin{document}
\begin{refsection}

\twocolumn[
  \begin{@twocolumnfalse}
\maketitle
\noindent
{\small
	\footnotesize{\textsf{$^1$ Cavendish Laboratory, Department of Physics, University of Cambridge, Cambridge CB3 0HE, United Kingdom}}\\
	\footnotesize{\textsf{$^2$ III. Physikalisches Institut, Georg-August-Universität, 37077 Göttingen, Germany}}\\
	\footnotesize{\textsf{$^3$ Department of Biochemistry, University of Cambridge, Hopkins Building, Cambridge CB2 1QW, United Kingdom}}\\
	\footnotesize{\textsf{$^4$ Adolphe Merkle Institute, Rue des Verdiers 4, 1700 Fribourg, Switzerland}}\\
	\footnotesize{\textsf{$*$ Corresponding authors: ullrich.steiner@unifr.ch, ch26@cam.ac.uk}}
}

\begin{abstract}
\noindent
\textsf{We tested the enhancement of electrical current generated from photosynthetically active bacteria by use of electrodes with porosity on the nano- and micrometer length-scale. For two cyanobacteria on structured indium-tin-oxide electrodes, current generation was increased by two orders of magnitude and the photo-response was substantially faster compared to non-porous anodes. These properties highlight porosity as an important design strategy for electrochemical bio-interfaces. The role of porosity on different length scales was studied systematically which revealed that the main performance enhancement was caused by the increased surface area of the electrodes. More complex microstructured architectures which spanned biofilms as translucent 3D scaffolds provided additional advantage in the presence of microbial direct electron transfer (DET). The absence of a clear DET contribution in both studied cyanobacteria, \textit{Synechocystis} and \textit{Nostoc}, raises questions about the role of conductive cellular components previously found in both organisms.}
\end{abstract}

 \end{@twocolumnfalse} \vspace{1em}
]

%

\noindent
Several microorganisms are able to generate electrons that can be collected and utilised in external circuits \autocite{McCormick2015}. In such devices, bio-anodes are the electrodes that collect electrons from the living bio-catalyst.
Bio-anodes in the best studied bioelectrochemical technology, microbial fuel cells, are commonly carbon or metal based, and a large diversity of morphologies has been used \autocite{Wei2011}. The electrode porosity usually has a strong effect on device efficiency~\autocite{Wei2011}, but the associated change in volume, surface area, and organism contact area can rarely be disentangled from the variation of materials themselves that are used to achieve the different morphologies. These complicated correlations currently limit the understanding of design rules for electrochemical bio-interfaces.
Furthermore, because of a lack of transparency of most anodes, there has been little work on the benefits of using porous electrodes in microbe-based devices that rely on light absorption, referred to as `biophotovoltaics', except for one study using larger, eukaryotic, algal cells~\autocite{Thorne2011}. 
While photosynthetic microorganisms are expected to operate with a quantum efficiency of five to ten percent internally, electrode interfaces and microbial electron export pathways currently limit device efficiencies to much lower values \autocite{McCormick2015}.

In this study we tested the effect of electrode porosity at different length scales on the performance of bioelectrochemical devices. To achieve this goal, we have compared three different electrode morphologies of the same translucent material. Two photosynthetic microorganisms \emph{Nostoc punctiforme} and \emph{Synechocystis} sp.\ PCC 6803  were each placed on (\textit{i}) a non-porous indium tin oxide (ITO) electrode, (\textit{ii}) a thick `nanoporous' ITO nanoparticle film, and (\textit{iii}) a `microporous' inverse-opal structure made from the same nanoparticles, and their photocatalytic current generation was investigated. 

Doped metal oxides are popular transparent electrode materials for a wide range of electronic applications.  ITO is one of the best performing transparent electrode materials, as it has a large optical bandgap, making it transparent to visible light, while the high levels of tin doping cause a metal-like conductivity. ITO is commonly used as thin film (tens of nanometres thick) in display applications and was shown to be biocompatible \autocite{Selvakumaran2008,Bombelli2012,Dewi2014}. ITO can be structured using a templating approach and has previously been used as porous glass in electrochemical studies of enzymes \autocite{Sokol2016,Noji2016}. In order to distinguish between the porosity on different length scales, we define nanoporosity as the presence of pores between sintered nanoparticles (10--100\,nm), and microporosity as the pores created by microsphere templates (10--40\,\textmu m). The templated inverse-opal pores used in this study are unusually large and the electrodes are unusually thick (80--140\,\textmu m) in order to accommodate a sufficient number of microorganisms within the structure to absorb incoming light. 

The two porosity length scales were chosen to represent biologically relevant regimes of electron transfer from microorganisms to the anode. The extracellular electron export mechanisms in cyanobacteria are still unclear, even for the model organism \emph{Synechocystis} \autocite{McCormick2015,Lea-Smith2016}.
Research on microbial electrochemical devices distinguishes between direct electron transfer (DET) from microorganisms to electrodes, and mediated electron transfer (MET) facilitated by electrochemically active molecules in solution. Nanopores are not directly accessible by the relatively large microbes but the increased surface area is available for electrochemical interactions with redox-molecules in the aqueous electrolyte, which is relevant for MET pathways. Pore sizes comparable to the cell size in microporous morphologies  allow cells to enter the electrode, thereby providing a considerable increase in direct contact area between bacteria and electrode surface.

\begin{figure}[tbp]
	\centering
	\includegraphics[width=\linewidth]{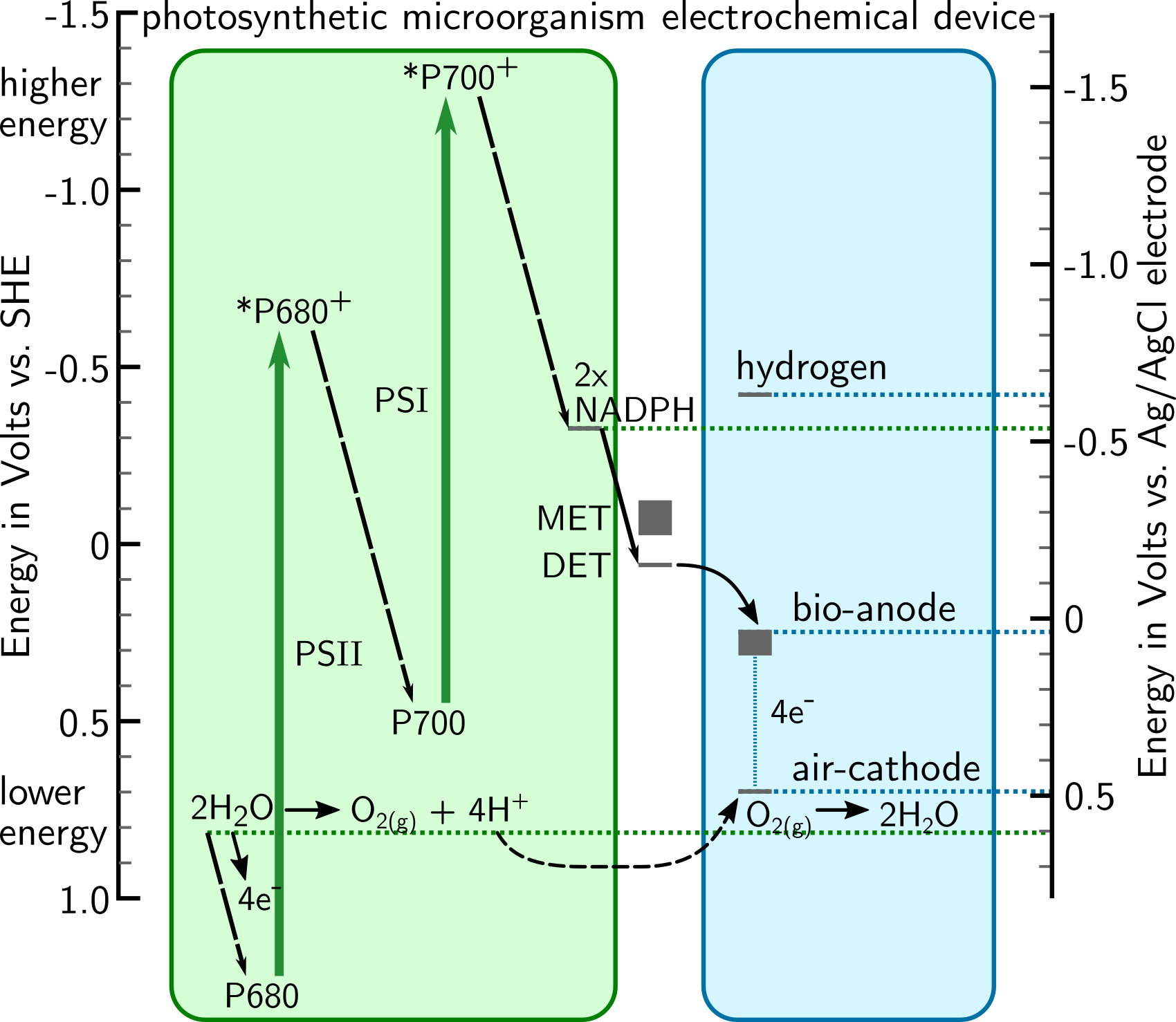}
	\caption{{\bf BPV energetic scheme.} Energy levels inside photosynthetic organisms and charge transfer to external electrodes of biophotovoltaic devices.}
	\label{FIGmainScheme}
\end{figure}

Biophotovoltaic devices generate electric energy though photo-electrons generated by living photosynthetic organisms on the anode, and a cathodic reaction taking place at a lower potential, here recombination to water. The energy levels involved in this process are detailed in the schematic diagram in Fig.\,\ref{FIGmainScheme}. The starting point of photosynthesis is photocatalytic water splitting, during which the electrons are gained. This is achieved through the absorption of two photons by the two photosystems (PSII and PSI), which form the characteristic Z-Scheme. The derived electrons may be stored through the production of NADPH molecules and can leave the cells via extracellular electron transfer (MET \& DET). The literature suggests similar values for redox-potentials of electron export in known organisms~\autocite{Schroeder2007} \ins{but with different efficiencies}~\autocite{Schuergers2017}. Taking interfacial energy transfer losses into account, these potentials determine the maximum voltage that can be generated in the device, versus the fixed potential of a platinum air-cathode. In practice, the cathode potential of air-cathodes does not reach the theoretical level of water splitting~\autocite{Schroeder2007}, which is indicated by the higher position of the cathode level in the scheme.

We used two different cyanobacteria in this study, \emph{Nostoc punctiforme} and the model species \emph{Synechocystis} sp.\ PCC 6803. With either of these microorganisms as bio-catalyst, both porous bio-anodes exhibited a remarkable ca.\,300 fold increase in generated peak photocurrent compared to non-porous ITO films that are routinely used for biophotovoltaic devices. We also studied the effects of the different anodes on the non-photosynthetic microorganisms \emph{Shewanella oneidensis} and found similarly dramatic increases in external current generation, with an additional enhancement on microporous electrodes.

\section*{\textsf{Results}}

	\begin{figure*}[tbp]
		\centering
		\includegraphics[width=\textwidth]{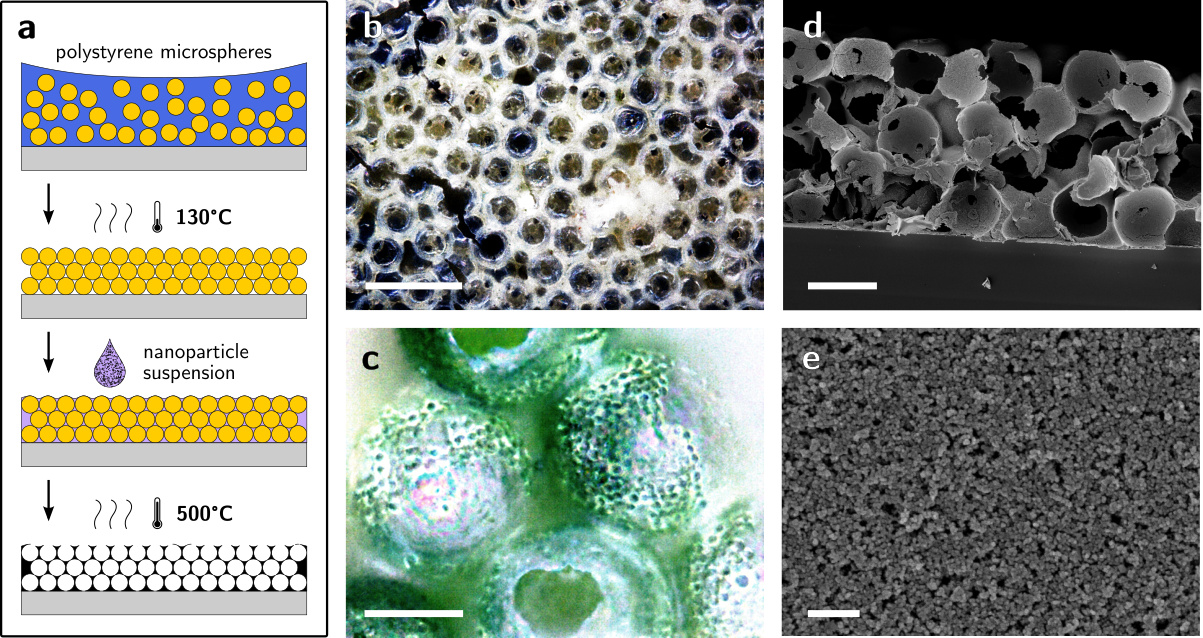}
		\caption{{\bf Electrode porosity}. ({\bf a}) Scheme of microporous electrode fabrication method. ({\bf b}) Optical bright field microscopy image of an empty microporous electrode (scale bar: 100\,\textmu m). ({\bf c}) Optical dark field microscopy image showing an electrode filled with cyanobacteria (after rinsing with water; scale bar:  20\,\textmu m). Scanning electron microscopy (SEM) images showing ({\bf d}) the cross-section of the electrode (scale bar:  40\,\textmu m) and ({\bf e}) the nano-porous structure  of sintered nanoparticles (scale bar: 400\,nm). }
		\label{FIGmainSEM}
	\end{figure*}
	
\paragraph{Electrode design and characterisation.}
A porous electrode that can incorporate photo-active biofilms must combine the three qualities (\textit{i}) conductivity, (\textit{ii}) translucency and (\textit{iii}) micro-porosity on a length scale that allows cells to enter interconnected pores while also forming biofilms with a thickness up to millimetres. To this end, an inverse opal structure was designed with pores and pore-connections of several micrometers, similar to that of cyanobacterial cells. The inverse opal structure was generated through a templating approach (Fig.\,\ref{FIGmainSEM}a). Polymer micro-spheres of 40\,\textmu m in diameter were deposited to form an opal structure, which was annealed to promote sphere  adhesion and to control the diameter  of the sphere-sphere interconnects, followed by  infiltration by a ITO nanoparticle suspension via capillary forces. Filling the template with nanoparticles ($<100$\,nm) proved to be the fastest and most reliable method to obtain thick porous films. A final heating step sintered the nanoparticles while burning out the polymer template to leave the inverse structure behind.

The resulting material was diffuse white-yellow, as seen by eye or optical microscopy (Fig.\,\ref{FIGmainSEM}b). After the addition of cyanobacterial cells, the electrode was examined by light microscopy (Fig.\,\ref{FIGmainSEM}c), to confirm the ability of the cells to populate the structure across the entire thickness (here, ca.\ 0.12\,mm). 
Cells (ca.\ 2\,\textmu m cell-diameter) had to pass through ca.\ 10\,\textmu m wide connections  in order to reach all hollow spheres of the inverse opal morphology (see Fig.\,\ref{FIGmainSEM}b and cross-section in d) or reach these spaces via cracks. The nano sized pores of the electrode material (Fig.\,\ref{FIGmainSEM}e) could not be accessed by the bacteria.

\begin{figure}[H]
	\centering
	\includegraphics[width=0.75\textwidth]{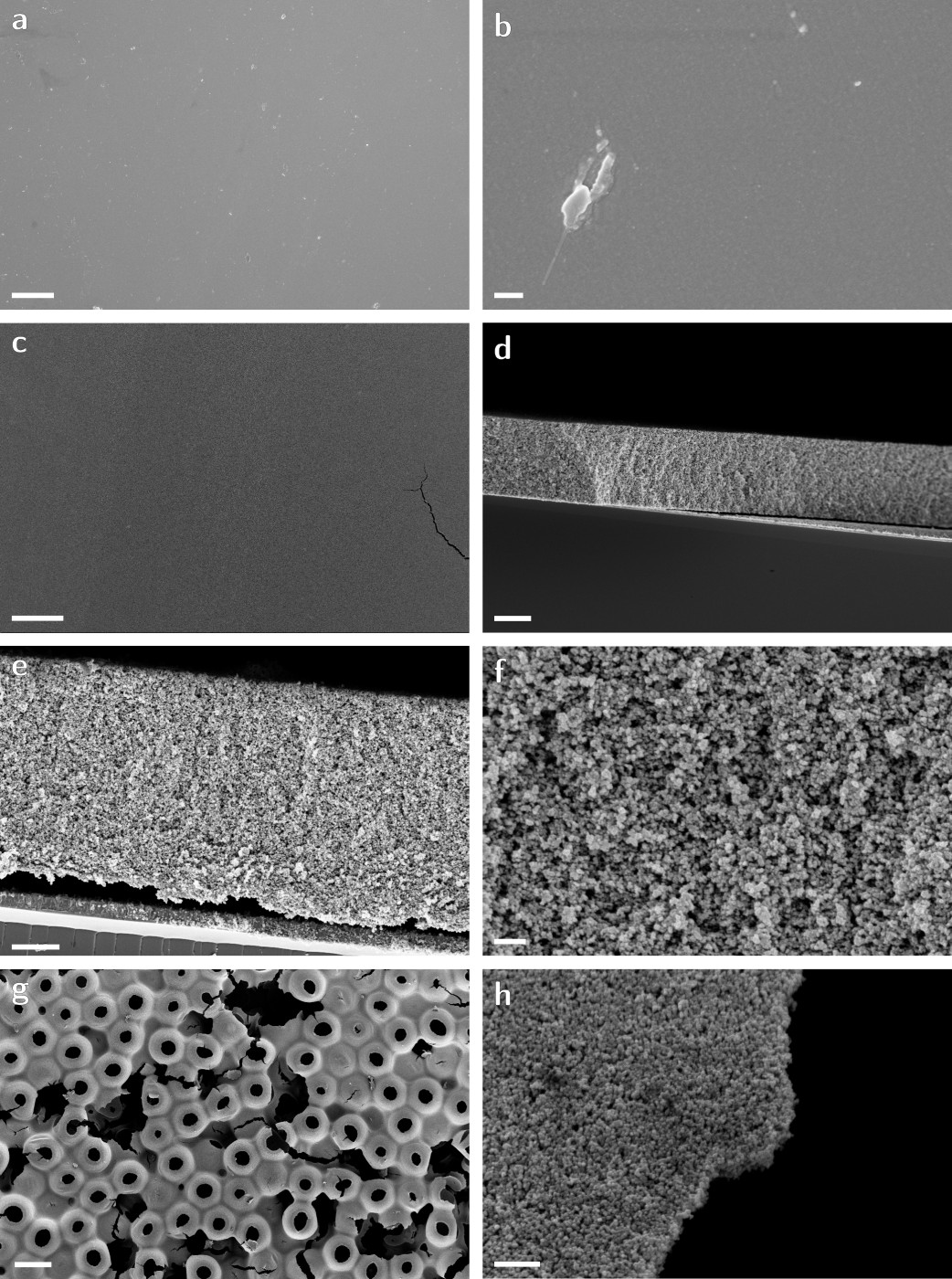}
	\caption{SEM images of the three anode types. (a,b) Commercial Ínon-porous ITO film on PET, with scale bars of 40\,\textmu m and 400 nm, respectively. (c) Nanoporous film of sintered ITO nanoparticles, top view, with scale bar of 40\,\textmu m. (d,e,f) Cross-section of the nanoporous electrode, scale bars are 4\,\textmu m, 2\,\textmu m, and 400 nm, respectively; (e) also shows the underlying conducting FTO layer on the glass substrate. (g,h) Top-view of  a microporous electrode, with scale bars of 40\,\textmu m and 500 nm respectively.  The defects in (b) and (c) are non-representative and  are shown to visualise the smoothness of the surrounding film.}
	\label{fig:subfigSEM}
\end{figure}

To compare the effects of micro-, nano- and no porosity in this study, three different ITO structures were used. A commercial non-porous ITO layer on a PET substrate served as reference (Fig.\,\ref{fig:subfigSEM}a,b), and thick nanoparticle films (Fig.\,\ref{FIGmainSEM}e, Fig.\,\ref{fig:subfigSEM}c-f) without and with additional micropores (Fig.\,\ref{FIGmainSEM}d, Fig.\,\ref{fig:subfigSEM}g,h) were employed to assess the interplay of nano- and microporosity on the charge generation by the microorganisms.

\begin{figure}[H]
	\centering
	\includegraphics[width=0.6\textwidth]{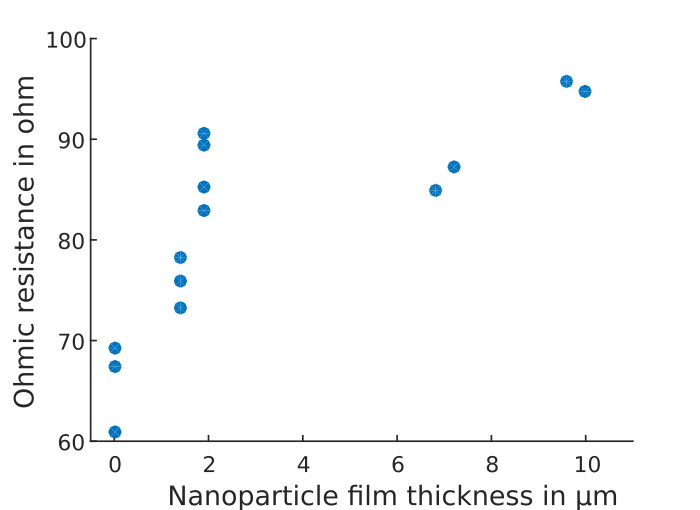}
	\caption{Resistance of different ITO nanoparticle film thicknesses. Data points represent the ohmic resistance (averaged real part of impedance measurement results) of dry nanoporous ITO `sandwich' samples on an ITO glass substrates, for different film thicknesses. Zero thickness refers to a pair of ITO substrates without a sandwiched coating.
	\rem{Four samples containing a 130-150\,\textmu m thick microporous ITO layers made from the same nanoparticles had resistance values of $110-165\,\Omega$}}
	\label{fig:subfigImpedance}
\end{figure}

The nanoparticle film had a thickness of ca.\ 9\,\textmu m and was produced with the same nanoparticles and on the same conducting FTO-glass substrate as the microporous electrodes. Its sheet resistivity on a non-conducting substrate was ca.\ 100\,$\Omega$/cm$^2$, determined by a 4-point probe.
\rem{To estimate the resistivity across the films as a function of film thickness and to compare it to that of thicker microporous electrodes, which contain narrow connection elements in its 3D-structure, impedance spectroscopy on dry electrode-`sandwich'-samples with an area of approximately 0.25\,cm$^2$ was additionally performed (see Methods for further detail).}
\ins{The film resistivity was determined by impedance spectroscopy (see Methods). The resistivity of dry electrode-`sandwich'-samples with an area of approximately 0.25\,cm$^2$ were measured as a function of film thickness. The obtained values were then compared to that of thicker inverse opal structured electrodes, the conductivity of which may be limited by their 3D morphology.} 
The electrodes exhibited purely ohmic resistance of 65 and 95\,$\Omega$ for temperature annealed ITO glass and 10\,\textmu m thick sandwiched nanoparticle films, respectively (Fig.\,\ref{fig:subfigImpedance}). The ca.\ 140\,\textmu m thick microporous electrodes displayed similarly low purely ohmic resistance of ca.\ 115\,$\Omega$.

The visual appearance and optical transmission spectra of the microporous and the nanoparticle film are shown in Fig.\,\ref{fig:subfigOptics}. Ca.\ 50\,\% of the visible spectrum was diffusely transmitted through the dry microporous film, and ca.\ 80\,\% through the dry nanoparticle film. Non-porous ITO is fully transparent to visible light, but causes partial specular reflection at air or water interfaces due to its high refractive index. Nanoporous ITO (Fig.\,\ref{FIGmainSEM}e) contains many internal interfaces that scatter the light, rendering electrodes translucent, which is even stronger for the inverse opal structure (Fig.\,\ref{FIGmainSEM}d). The relatively large opal unit cell size of 40\,\textmu m was chosen to minimise the number of metal oxide water interfaces and thus the back-scattering of light, and to allow the use of larger microorganisms. Scattering and absorption were more strongly pronounced in this electrode than in previously described ITO-based inverse-opal structures (e.g.\ in \autocite{Liu2014} ca.\ 80\,\% of the visible spectrum is transmitted), due to the unusually large thickness of the electrode.

\begin{figure}[tbp]
	\centering
	\includegraphics[width=\linewidth]{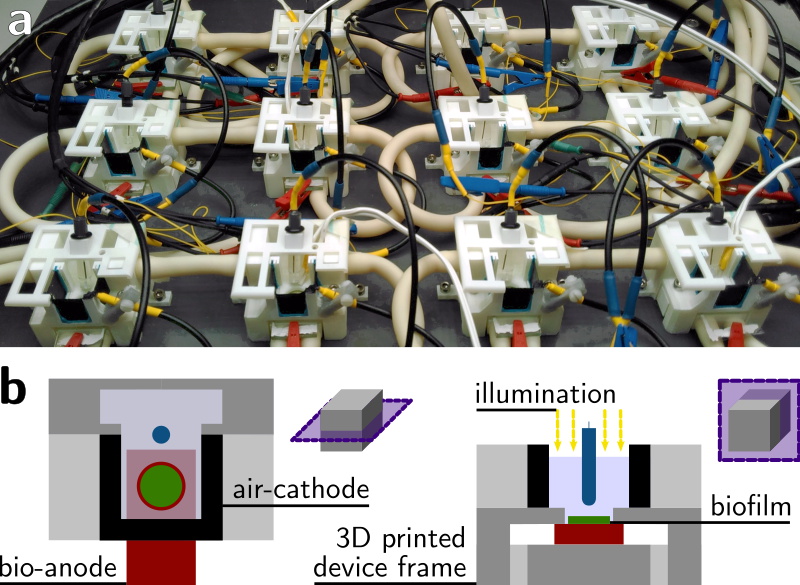}
	\caption{{\bf  Design of biophotovoltaic devices.} ({\bf a}) Assembled and connected 3D-printed devices, incl.\ reference electrodes with with blue connectors, positioned in the homogeneous illumination area.
	({\bf b}) Two schematic perspectives of cuts through a device indicating the locations of electrodes and biofilm.}
	\label{FIGmainDevice}
\end{figure}

\paragraph{Biophotovoltaic devices.}
Figure\,\ref{FIGmainDevice} pictures the electrochemical devices used for photo electricity measurements in up to 12 parallel channels. The devices were 3D printed to allow for a compact, shareable design and to enable temperature control via hollow walls and a circulating water bath. The structured anodes on FTO-glass were inserted from the bottom on a printed anode holder, and sealed with an O-ring. \ins{The devices were designed to leave a (macroscopic) electrode area of 1\,cm$^2$ exposed to microorganisms and light.} Three pieces of platinum nanoparticle based air cathodes with a combined area of ca.\ 6\,cm$^2$ were arranged upright and close to each anode opening. Parallel measurements with the same batch of bacteria culture and under the same high-intensity light source were essential to obtain reliable mean results in experiments.

\begin{figure*}[tbp]
	\centering
	\includegraphics[width=\textwidth]{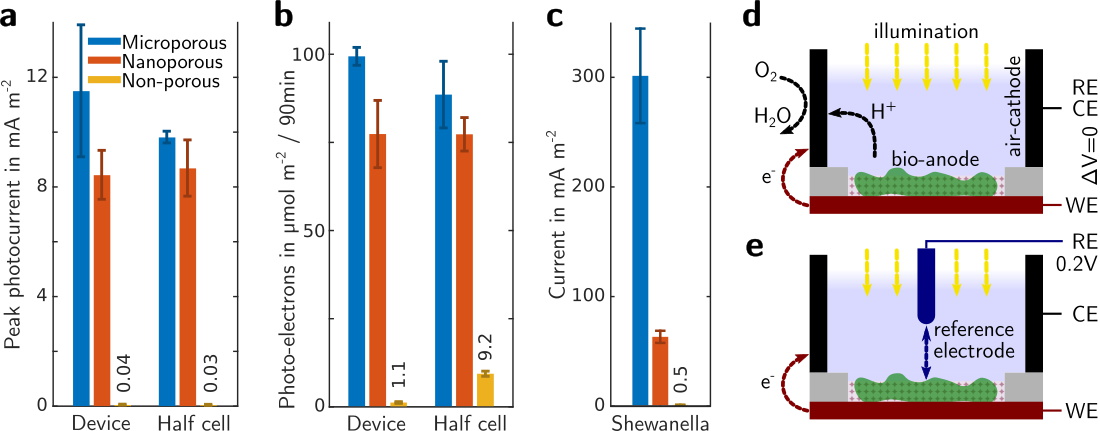}
	\caption{{\bf Current levels by anode type and device configuration.} ({\bf a}) \emph{Synechocystis} peak photocurrents in device mode (at `short-circuit') and half-cell operation (with reference electrode), during a photon flux of 460\,\textmu m/m$^2$/s. ({\bf b}) As (a), number of collected  charges during 90 minutes of illumination at photon flux of 512\,\textmu m/m$^2$/s. {\bf (c)} Current generated by Shewanella cells for each anode type. ({\bf d,e}) Schematic illustrating the device and half-cell operation, respectively. The potentiostat connections for reference (RE), working (WE) and counter electrode (CE) are indicated.}
	\label{FIGmainCurrentPeaks}
\end{figure*}

\paragraph{Photocurrent ratios.}

\begin{figure}[H]
	\centering
	\includegraphics[width=0.8\textwidth]{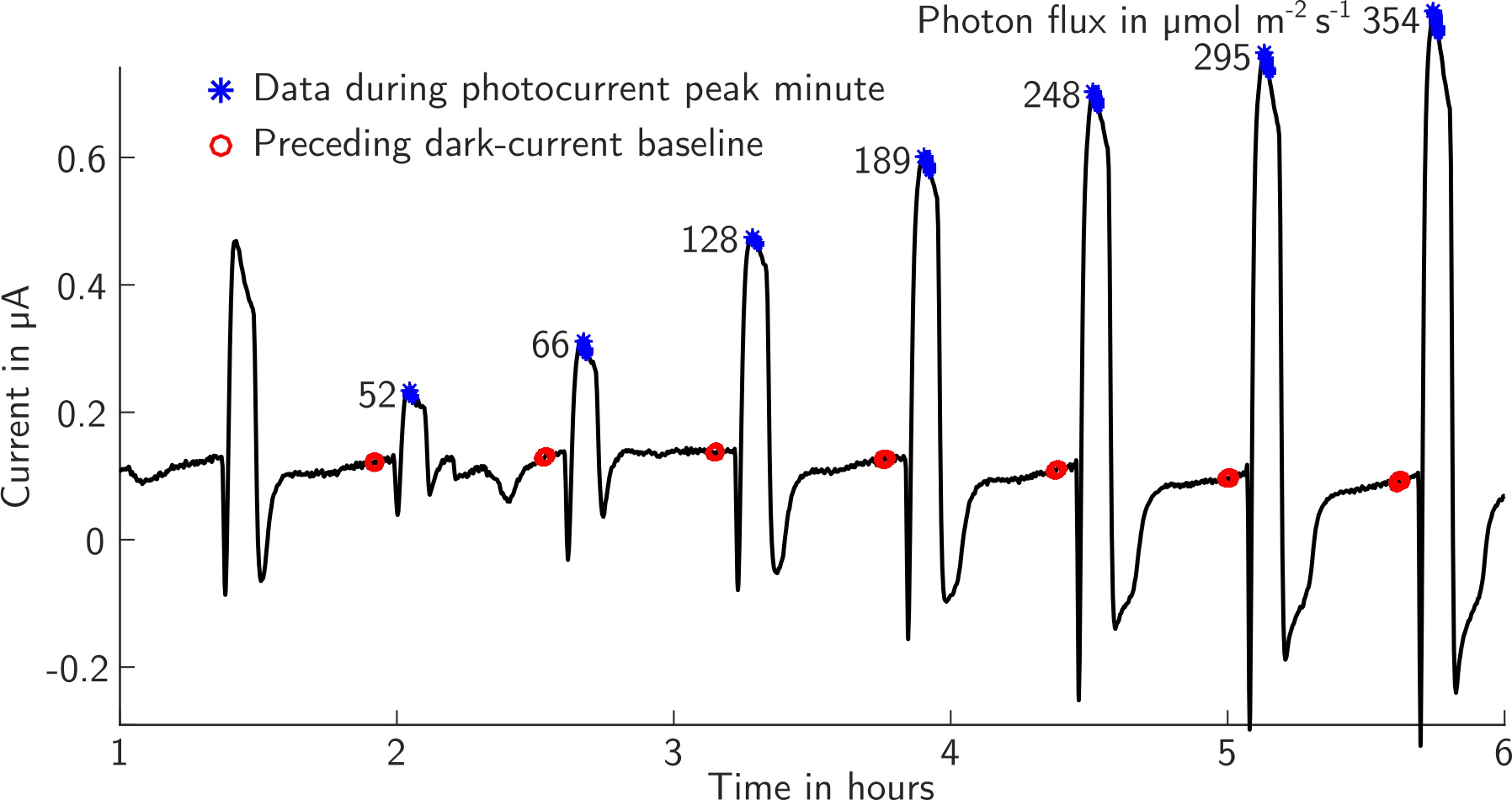}
	\caption{Visual explanation of peak currents. The first seven photocurrent peaks are shown of an illumination sweep with increasing light intensity\rem{(the initial dark current was subtracted)}. The photocurrent peaks are indicated with stars and numbers, along with data points during the peak minute. Averaging  peak-minute values provided a reliable value that was less dependent on noise and peak shape. Illumination periods where chosen short enough (5-7\,min.) and separated enough (20-30\,min.) to let the baseline dark-current recover before the next illumination. To account for baseline fluctuations, the magnitude of each photocurrent peak was measured from the preceding dark-current, here indicated by red circles. The base level was also taken as average of data during one minute. Underlying data is a chronoamperometry measurement fragment of a \emph{Synechocystis} BPV with nanoporous electrode operated at short-circuit.}
	\label{fig:subfigPeakIllustration}
\end{figure}

To study the effect of electrode porosity on the biophotovoltaic device performance, parallel measurements with equal numbers of cells (based on amount of chlorophyll) were conducted on the three different electrode structures. 
The peak photocurrent levels observed with \emph{Synechocystis} are shown in Fig.\,\ref{FIGmainCurrentPeaks}a.
\rem{Since the current did not stay constant during light exposure, the}
\ins{During the exposure of devices to light, the photo-current did not stay constant, but reached a peak value, followed by a decreasing slope, slowly reaching a steady value. The} peak photocurrent was defined as the average short-circuit current (0\,V external potential between anode and cathode and no reference electrode) measured during one minute around its peak value, minus the dark-current level before each illumination, see Fig.\,\ref{fig:subfigPeakIllustration}.

Both micro- and nanoporous bio-anode based devices substantially outperformed the non-porous ITO electrode with 11.5 and 8.4\,mA/m$^2$ to 0.04\,mA/m$^2$ respectively, while the microporous structure displayed a small but consistent advantage over the film with nanopores only. During the illumination period of seven minutes, the microporous devices reached a peak photocurrent value $\sim$300 times higher than those equipped with non-porous electrodes.

In half-cell operation, the total  charges collected during a longer period of 90 minutes were larger by more than a factor of 10 with microporous vs.\ non-porous anodes, see Fig.\,\ref{FIGmainCurrentPeaks}b.
Equilibrium values of the photocurrent are difficult to obtain because BPV generated currents tend to fluctuate over time. Photocurrent measurements are also limited by photo bleaching of photosynthetic pigments during illumination and the long time periods necessary for bio-anodes to stabilise after a change of conditions. To quantify the photovoltaic output of the devices, the number of electrons generated during 90 minutes of illumination (Fig.\,\ref{FIGmainCurrentPeaks}b) was therefore considered as a robust current measure in addition to the recorded peak values (Fig.\,\ref{FIGmainCurrentPeaks}a).

Similar results were obtained for \emph{Nostoc} bio-anodes, where the current enhancement of devices with a microporous electrode was also higher by a factor of $\sim$300 compared to non-porous electrodes, with a peak-current of 11.2\,mA/m$^2$, and 30 times more charge collected during 90 minutes (167\,\textmu mol/m$^2$).

Figure\,\ref{FIGmainCurrentPeaks}d illustrates the BPV device chamber at `short-circuit' in which the cathodic water recombination reaction drives the device current and voltage, without an additional force of an externally applied potential.
Electrochemical studies commonly use a half-cell configuration instead (Fig.\,\ref{FIGmainCurrentPeaks}e). There, the anode potential is set with respect to a reference electrode of a stable and known potential, here 0.2\,V vs.\ an Ag/AgCl reference electrode, and the cathode is dynamically shifted to a potential where the cathodic reaction is non-limiting for the measurement. To test whether the cathode limits the peak photocurrents reached in our devices, a set of peak and continuum measurements were performed in half-cell mode, shown alongside the device measurements in Fig.\,\ref{FIGmainCurrentPeaks}a,b. 
The highest current levels observed, here in the case of porous anodes, were consistent between device and half-cell operation within each others standard deviations, indicating the absence of cathode limitation.

For comparison with an organism of well-studied ability to perform direct electron transfer \autocite{Gorby2006}, an additional test was performed with non-photosynthetic \emph{Shewanella oneidensis} bacteria in the three electrode set-up. The average generated currents  were 0.5\,mA/m$^2$, 62\,mA/m$^2$, and 299\,mA/m$^2$ for non-porous, nanoporous, and microporous bio-anodes, respectively (Fig.\,\ref{FIGmainCurrentPeaks}c), recorded in half-cell mode at an anode potential of 0.2\,V vs.\ a Ag/AgCl reference electrode in fresh LB medium. This corresponded to a more than 100-fold current increase from non-porous to nanoporous electrodes, and a further ca.\ 5 times increase for the microporous structure.

\begin{figure}[tbp]
	\centering
	\includegraphics[width=\linewidth]{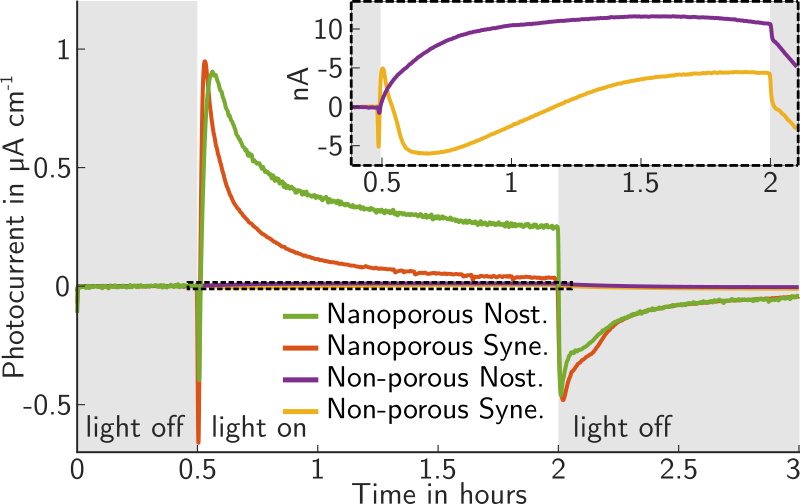}
	\caption{{\bf Light response dynamics.} Development of the average BPV photocurrent over time at a photon flux density of 512\,\textmu mol/m$^2$/s, for non-porous and nanoporous film electrodes and both \emph{Nostoc} and \emph{Synechocystis}.}
	\label{FIGmainCurrentDynamics}  
\end{figure}

\begin{figure}[H]
	\centering
	\includegraphics[width=0.7\textwidth]{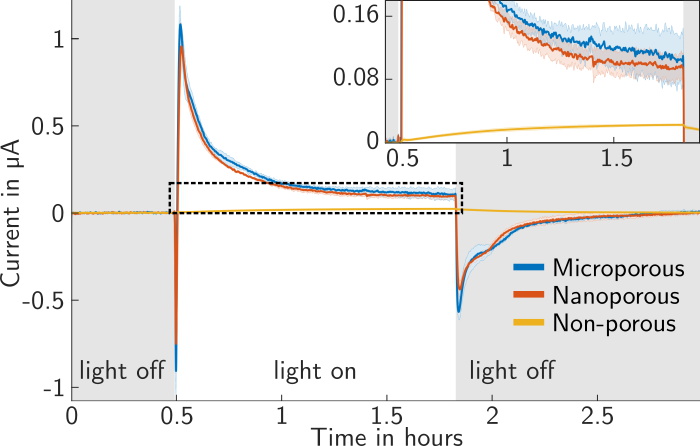}
	\caption{Light response characteristics of the three different \emph{Synechocystis} bio-anode types over time, operated in half-cell mode at 0.2\,V vs.\ an Ag/AgCl reference electrode. The photon flux during illumination was 512\,\textmu m/m$^2$/s. Averages and standard deviations (shaded) where formed from 3-4 individual devices per anode type.}
	\label{fig:subfigChronoampEq}
\end{figure}

\paragraph{Light response characteristics.}
Exposing photosynthetic electro-active biofilms to light gave rise to an initial current peak which typically dropped to steady-state values. For each of the organisms studied here, the photocurrent rose much faster to its peak value on devices employing one of the porous compared to non-porous electrodes, with little difference between the two pore types (Fig.\,\ref{FIGmainCurrentDynamics} and \ref{fig:subfigChronoampEq}). The current peaks were reached for porous electrodes after 1--6 minutes, whereas devices with non-porous electrodes required up to over one hour of light exposure to reach a maximum (Fig.\,\ref{FIGmainCurrentDynamics} and Fig.\,\ref{fig:subfigChronoampEq} expanded regions). The steady-state current reached by non-porous bio-anodes was still lower compared to the porous electrodes.
\ins{Note that the photo-current minimum for \emph{Synechocystis} cells on a non-porous electrode (Fig.\,\ref{FIGmainCurrentDynamics}) is unexpected.  It was not observed when the measurement was carried out with reference electrode (Fig.\,\ref{fig:subfigChronoampEq}). Since this study is concerned with the current maximum which was reached after two hours, the origin of the minimum was not further investigated.}

\begin{figure}[tbp]
	\centering
	\includegraphics[width=\linewidth]{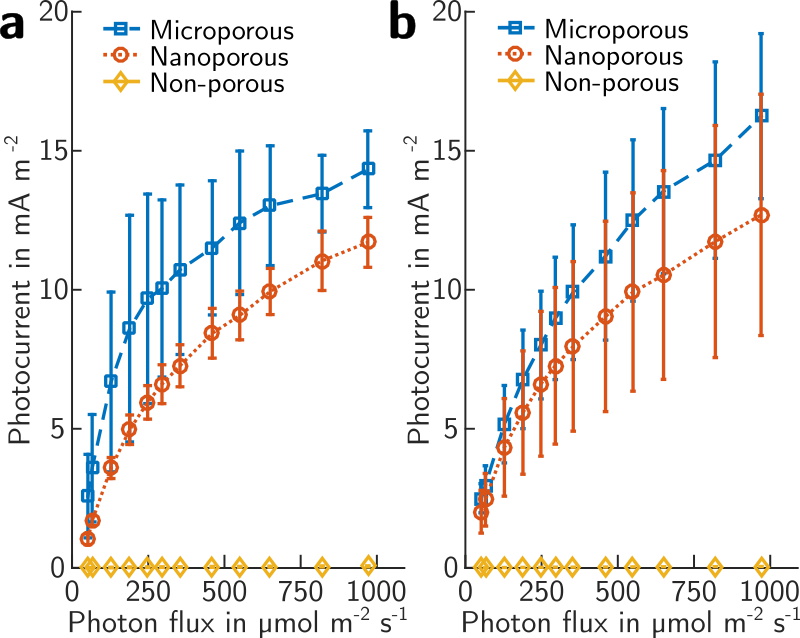}
	\caption{{\bf Saturation behaviour.} ({\bf a}) Peak current generated by \emph{Synechocystis} cells on the three anode types at different light levels. The values are averages of the peak minute after subtracting the dark current. ({\bf b}) \emph{Nostoc} biofilm peak photocurrents as in (a).}
	\label{FIGmainSweeps}  
\end{figure}

The fast photo response of the porous bio-anodes enabled serial experiments studying  the peak BPV photocurrent for different light irradiances, see Fig.\,\ref{FIGmainSweeps}a,b. The resulting curves show an approximately exponential saturation of photoelectron generation, with higher efficiencies \ins{(ratio of photocurrent to photon flux)}  reached at low light levels. 
The performance advantage of porous electrodes persisted for all irradiation levels. 
\ins{The similarity of photocurrents on porous electrode types when used with either \emph{Synechocystis} or \emph{Nostoc} indicates the importance of the effective mesoscopic electrode surface area rather than its coarse morphology. At the same time, the similar saturation curves for each of these morphologies confirms the expectation that the microporous electrode architecture does not shade cells to a significant degree.} 

In addition to the use of porous bio-anodes, the need for fast and accurate measurements was addressed (\textit{i)} by adding phosphate buffer to the BG11 medium which increased the electrolyte conductivity, (\textit{ii)} by disregarding preceding dark-current levels, and (\textit{iii)} by choosing dark-times between (short) illuminations that were long enough for the dark current level to recover. For the low light level \emph{Nostoc} measurements (photon flux 0--400\,\textmu mol/m$^2$/s,  Fig.\,\ref{FIGmainSweeps}b), the irradiation intervals were not long enough to reach the full peak values, which may explain the lower exponential slope for low photon fluxes compared to the \emph{Synechocystis} data.

\begin{figure}[H]
	\centering
	\includegraphics[width=0.85\textwidth]{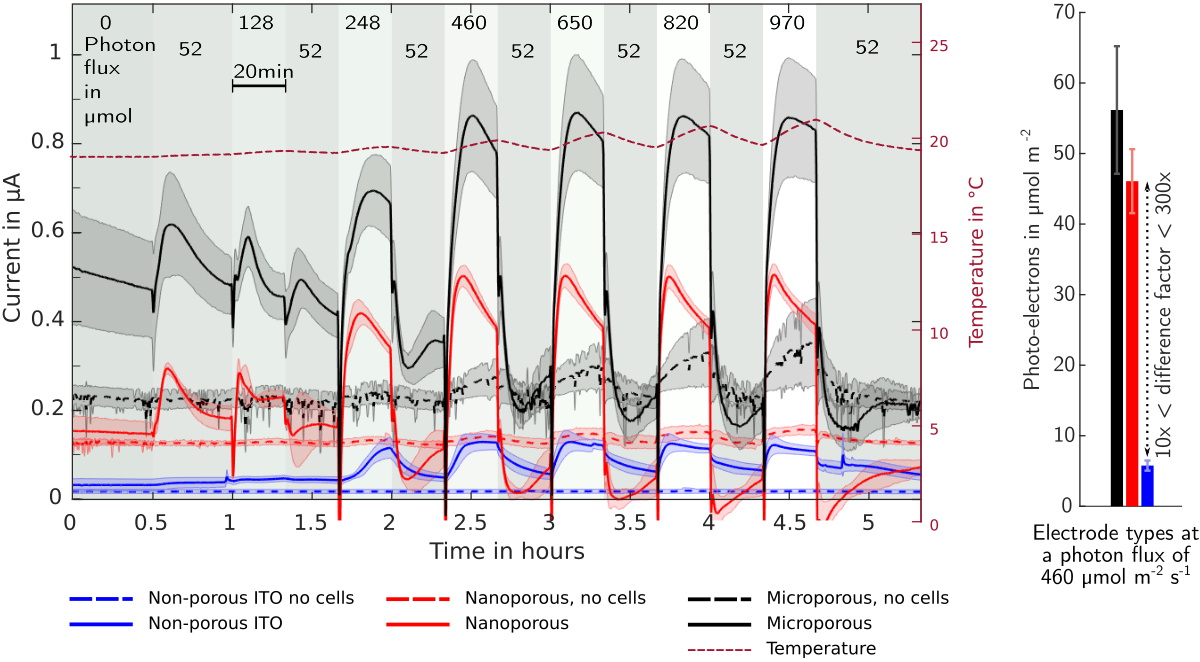}
	\caption{Current generated by electrochemical devices (in half cell mode at 0.2\,V vs.\  Ag/AgCl) with and without biofilms (\emph{Synechocystis}) operated at different light intensities. The electrolyte in the devices was  BG11 medium without supplementary phosphate buffer. The temperature curve corresponds to the values measured in the electrolyte close to the anode. In contrast to other measurements presented in this study, the dark-times between illumination times were not chosen to be long enough in this measurement for the dark current level to recover to its pre-illumination base-level.}
	\label{fig:subfigChronoSweepPrevious}
\end{figure}

Surprisingly, in contrast to porous electrodes, non-porous ITO electrodes performed better and responded faster in the absence of phosphate buffer, see Fig.\,\ref{fig:subfigChronoSweepPrevious}, but its performance remained much below that of the microporous electrodes, which collected ten times more electrons during a 20 minute illumination period.

\paragraph{Redox reactions at the bio-anode.}
\rem{The generation of electrons by bio-catalytic water splitting depends only on the input energy of the illuminating light source.}  
\rem{These independently produced electrons can however only be collected in a BPV device if the bio-anode potential is lower than the energy of electron export processes.}
\ins{The activity of electron-generation by bio-catalytic water splitting depends only on the input energy of the illuminating light source (Fig.\,\ref{FIGmainScheme}).}
By measuring the electron collection as a function of anode potential by cyclic voltammetry (CV)
\rem{typically results in an s-shaped curve characterised by a quick rise in current to a stable value at the potential of the electron export processes, and an equivalent current drop during the backwards scan.}
\ins{the redox activity of downstream electron donating molecules can be characterised}~\autocite{Harnisch2012}.
The bio-catalytic activity is slow for most organisms, however, and could not be detected clearly by CV even at scan rates as low as 0.5\,mV/s. The porous anodes showed a strong enhancement in electrochemical sensitivity, but also stronger surface charging, leading to high \rem{faradaic} \ins{charging} currents \ins{(non-Faradaic)} that broaden the CV hysteresis curve and thus mask bio-catalytic peaks, see Figs.\,\ref{fig:subfigFaradaicCurrent} and \ref{fig:CVtypesmedia}. \ins{This charge accumulation is also reflected by the transient spikes observed in chronoamperometry measurements when the light is turned off (e.g.\ Figs.\,\ref{FIGmainCurrentDynamics} and \ref{fig:subfigPeakIllustration}).}

\begin{figure}[tbp]
	\centering
	\includegraphics[width=\linewidth]{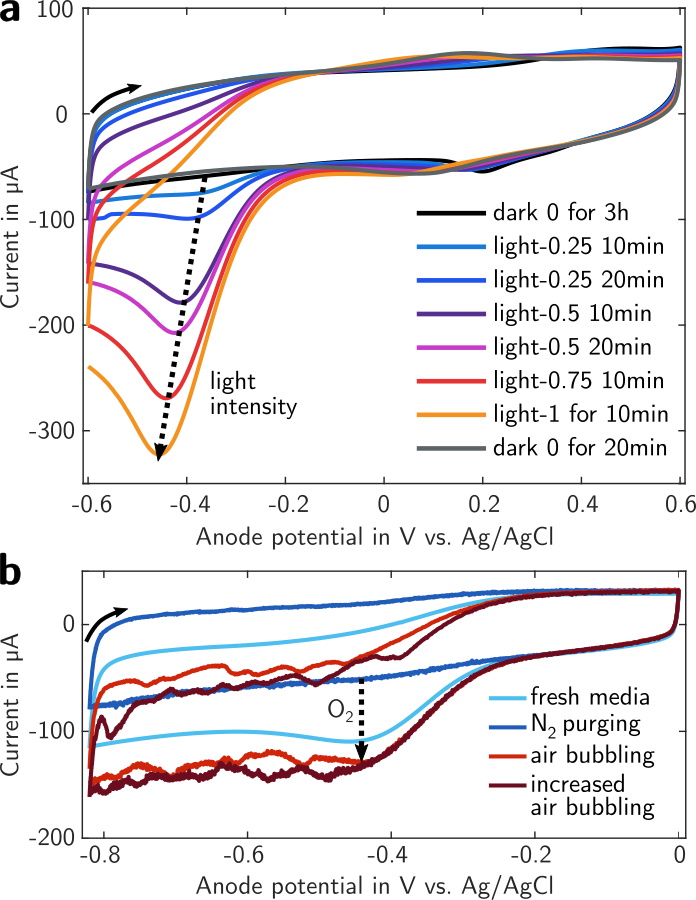}
	\caption{{\bf Cyclic voltammetry analysis} of microporous ITO anodes at a scan rate of 4\,mV/s with BG11 medium as electrolyte. ({\bf a}) Response of the bio-anode (\emph{Synechocystis}) under exposure to different constant light levels. Light intensity 1 corresponds to a photon flux of 820\,\textmu m/m$^2$/s, and the time indicates for how long the light level was applied by the end of the measurement. ({\bf b}) Scans of abiotic anodes in the dark. The oxygen content of the electrolyte was varied by purging the electrolyte with nitrogen gas or compressed air.}
	\label{FIGmainCV} 
\end{figure}

\begin{figure}[H]
	\centering
	\includegraphics[width=0.55\textwidth]{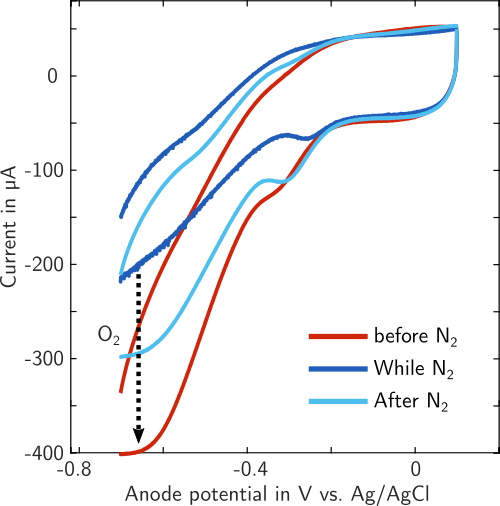}
	\caption{Cyclic voltammetry analysis of a microporous bio-anode with \emph{Synechocystis} cells at a scan rate of 4\,mV/s. Scans from -0.71 to 0.1\,V vs.\ Ag/AgCl at a photon flux density of 295\,\textmu m/m$^2$/s. The tracks were recorded before, while and after purging the electrolyte with nitrogen gas to remove the oxygen released by the photosynthesising organisms.}
	\label{fig:subfigCVpurged}
\end{figure}

The increased electrochemical sensitivity of microporous electrodes did however enable the detection of a light dependent  reversible peak of a reduction reaction on the anode surface (Fig.\,\ref{FIGmainCV}a). This peak was not  observed on the less sensitive non-porous ITO anodes. The slight shift in peak position with irradiance  was  probably caused by a pH shift of the BG11-medium electrolyte (no additional phosphate buffer was used in this measurement).
Lowering the oxygen content of the electrolyte by purging it with nitrogen gas led to a decrease in peak height (Fig.\,\ref{fig:subfigCVpurged}), suggesting  its link to photosynthetic oxygen production. This was further confirmed by a measurement without microorganisms, during which the oxygen content was varied by purging with nitrogen or air, and for which the same reduction peak was reduced (nitrogen) and enhanced (air), see Fig.\,\ref{FIGmainCV}b.
The reaction appeared at anode potentials lower than --300\,mV, and was well separated from the operating range of presented bio-anodes, which had a measured electrode potential of 50 to 150\,mV vs.\ Ag/AgCl (Fig.\,\ref{FIGmainScheme}). This separation is important for device performance, as the electron-consuming reduction of oxygen can compete with the collection of photosynthesis-derived electrons (oxidation reaction) if it takes place at the material- and pH-dependent operating potential of the bio-anode.

\section*{\textsf{Discussion}}

Translucent conductive electrodes with porosity on two relevant length scales -- nanoporous films with pores accessible only by electrolyte, and microporous electrodes with additional pores on the organism's length scale -- were successfully created. Two cyanobacteria,  \emph{Synechocystis} sp.\ PCC 6803 and \emph{Nostoc punctiforme}, were applied to these electrodes where they continued to be electroactive for the duration of measurements (at least several days), with no indication of bleaching. 
During measurements, oxygen reduction at the anode is unlikely to compete with the collection of microorganism-derived electrons, as it takes place at lower anode potentials (--400\,mV) than those reached during BPV operation (ca.\ 100\,mV).

In the architecture presented here, the cathode did not limit device performance (observed peak photocurrents), and half-cell measurements were equivalent to overall device performance. Measuring the bio-anode versus a reference electrode (half-cell mode) simplifies the interpretation of results of the anode performance and the presence of a reference potential allows the application of methods such as cyclic voltammetry. However, half-cell measurements can only be related to external quantum efficiencies as defined in the field of solar energy generation, when it is known that an appropriate non-limiting cathode such as the here-presented can be designed to complement it in a full device.

\begin{figure}[H]
	\centering
	\includegraphics[width=0.7\textwidth]{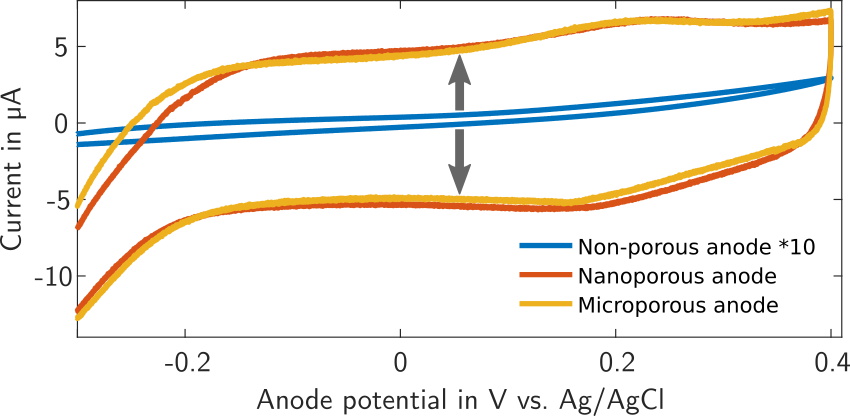}
	\caption{\ins{Non-Faradaic charging} current on porous electrodes broadens cyclic voltammetry (CV) scans. Example CV data at scan rate of 0.5\,mV/s on \emph{Synechocystis} bioanodes differing in porosity is shown. As indicated by arrows, data from scans performed on porous anodes are broadened in comparison to non-porous electrodes (the later was ten times enlarged for clarity) due to enhanced \rem{faradaic} \ins{charging} currents. They are also more sensitive to detect electrochemically relevant molecules near the electrode, as described in the results section.}
	\label{fig:subfigFaradaicCurrent}
\end{figure}

\begin{figure}[H]
	\centering
	\includegraphics[width=0.7\textwidth]{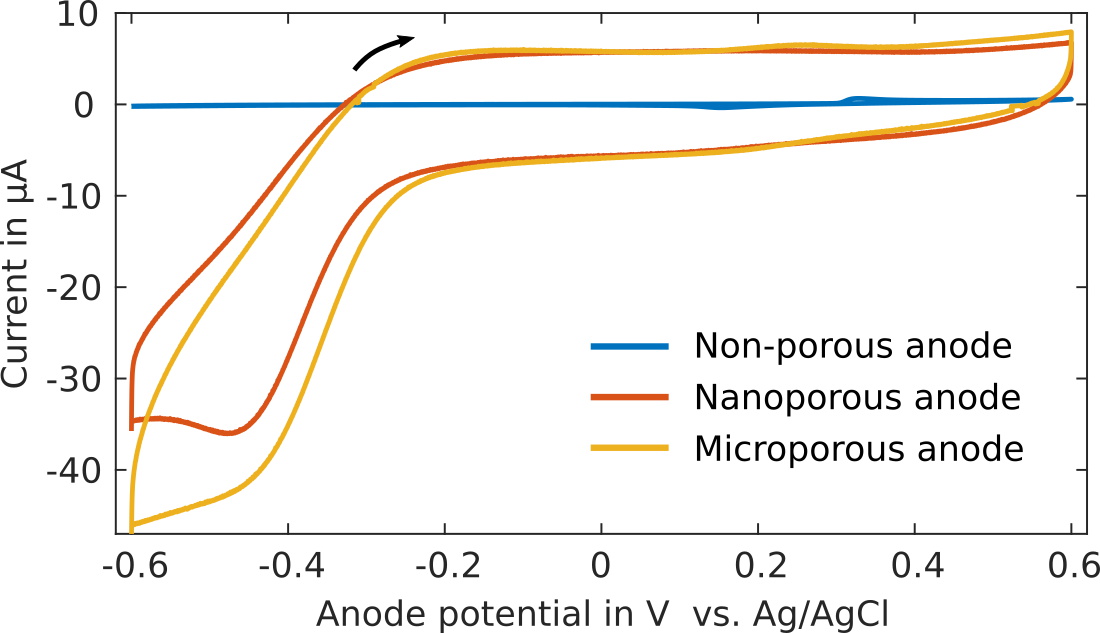}
	\caption{Data demonstrating the extend of surface charging by anode type. Individual cyclic voltammetry scans were taken at a scan rate of 0.5\,mV/s with BG11 medium (without microorganisms and light) on anodes differing in porosity. Charging currents were comparable for both porous anode types but much lower for non-porous anodes.}
	\label{fig:CVtypesmedia}
\end{figure}

Photocurrents collected from both types of porous electrodes were typically one to two orders of magnitude larger than those from bio-anodes based on non-porous ITO electrodes.  The performance advantage varied depending on the metric used (higher for peak currents; lower for charges collected over several hours) and electrolyte conductivity (higher salt levels improved the performance of porous anodes). Photocurrent peak values were also reached faster (up to 90 times) on porous electrodes. \ins{The similar photocurrent levels of both porous electrode types for each cyanobacteria species (Fig.\,\ref{FIGmainCurrentDynamics}) indicates that the current enhancement can be attributed to a non-limiting electrode surface area. The porous electrodes accordingly displayed similar coulombic currents caused by surface-charging  (Figs.}\,\ref{fig:subfigFaradaicCurrent} and \ref{fig:CVtypesmedia}). Both high performance and fast response demonstrate the importance of a large electrode surface area for the design of electrochemical bio-interfaces.

The fast response of porous bio-anodes enabled the performance of timely serial experiments measuring the dependence of photocurrents on light irradiation. The resulting curves show an approximately exponential saturation behaviour.
The correlation of this measurement to the saturated process of photosynthetic electron transfer itself may point to a direct link between the levels of photo-electrons available inside the cell, e.g.\ as the ratio of NADPH/NADP$^+$ molecules, and the peak number of photo-electrons measured in a BPV device.
The magnitude and temporal variation of photocurrent generation is not yet well understood, and would depend on a complex interplay of biochemical processes, including competition with the terminal oxidases, which act as electron sinks \autocite{Bradley2013}. Further research into the role of pigment concentration, saturation of light absorption and the regulation of terminal electron sinks within the cell will provide valuable insights into biophotovoltaics, as well as into the bioenergetics of photosynthesis in general. The quantitative datasets from the fast-response bio-anodes described here could provide an important contribution towards this goal.

The high electrochemical sensitivity of the porous anodes, particularly the microporous electrode enabling biofilm penetration, also enabled a sensitive real-time detection of the oxygen levels present inside photosynthetically active biofilms by cyclic voltammetry (Fig.\,\ref{FIGmainCV}). Such analyses might provide quantitative insights into oxygenation studies e.g.\ in mixed biofilms or tissues~\autocite{Schenck2015}.

Surprisingly, the performance of nano- and micro-porous electrodes was similar for both cyanobacteria. This similarity simplifies design rules of electrodes, requiring simply the sintering of a nanoparticle paste for the manufacture of nanoporous layers instead of more elaborate structuration. 
It also implies that the dominating extracellular electron transfer mechanism may be self-mediated (no additional electron shuttling molecules were added). If the cyanobacteria were able to inject electrons efficiently into the electrode surface via direct electron transfer (DET), a larger performance increase would be expected for the microporous electrode, because it provides a many-fold increase in organism-electrode contact area compared to the two films.
The potential DET-related performance increase was confirmed by using non-photosynthetic bacterium \emph{Shewanella oneidensis}, which is well known to exhibit DET \autocite{Gorby2006} and for which microporous bio-anodes showed a clear (ca.\,5-fold) additional advantage over nanoporous films. 
DET tends to be fast \rem{and low-loss} \ins{while avoiding the diffusion losses associated with soluble electron carriers}~\autocite{Schroeder2007} and can provide an important contribution to  currents  collected from electroactive biofilms. \rem{The most} So-called `nanowires'  \ins{have been proposed} to give rise to an efficient DET mechanism in some microbial biofilms. Nanowires are conductive extracellular pili-like structures that  \rem{have been shown to} transport redox-electrons from within the cell through the insulating extracellular matrix \ins{and neighbouring biofilm cells} to electrodes \autocite{Gorby2006, Malvankar2011} \ins{but their electrochemical properties are still disputed}~\autocite{Yates2016}. Conductive extracellular pili-like structures (PLS), ca.\,10\,\textmu m in length, have recently been identified in the cyanobacteria utilised in this study, \emph{Synechocystis} sp.\ PCC 6803 and \emph{Nostoc punctiforme}, as well as in the cyanobacterium \emph{Microcystis aeruginosa}, and it was speculated that they contribute to extracellular electron transfer in these organisms \autocite{Gorby2006, Sure2015, Sure2016}. 
According to this hypothesis, the filaments should lead to a strong enhancement in photocurrents when brought in contact with the electrode surface across the entire photosynthetic biofilm, by using the microporous structure. The performance similarity of biofilms on micro- and nanoporous electrodes however suggests that the potential presence of conductive bacterial filaments did not contribute significantly to the photo-electron export in our devices, because the filaments did not connect to the organism's photosynthetic electron transfer chain or to the electrode, or the filaments were not produced. 

The measurements presented here suggest a slight advantage of the microporous electrode morphology over the nanoporous film. 
This advantage might be due to a larger, more accessible electrode surface, a smaller average distance self-mediated small molecules have to travel to reach the electrode, or the fact that biofilms reach further into the microporous electrodes,which therefore probe the photoresponse of bacteria that are exposed to high irradiation levels and thus contain a larger number of available photo-electrons. On the other hand, the performance of  microporous electrodes  may be  negatively affected by increased scattering of the inverse opal structure, which leads to slightly increased reflection and decreased transmission, as well as an increase in electrode resistance with increasing thickness of the electrode.

It is clear that use of porous translucent electrodes offers a dramatic increase in the current density obtainable from photosynthetic microorganisms in biophotovoltaic devices. Their rapid photo-response times may additionally allow these devices to be exploited as a wider tool for the study of photosynthetic electron transfer.



\section*{\textsf{Methods}}
\begin{figure}[H]
	\centering
	\includegraphics[width=1\linewidth]{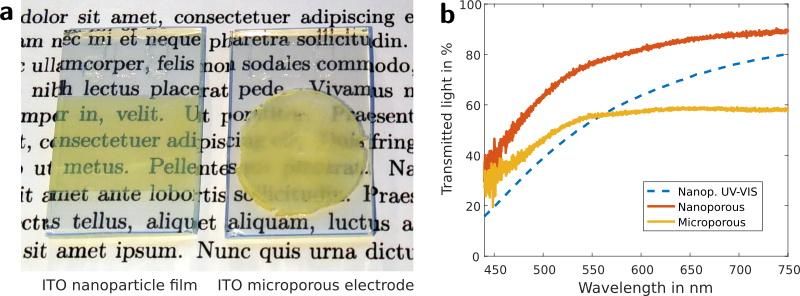}
	\caption{Macroscopic appearance of the translucent electrode material. (a,b) Photographs of  nano- and microporous ITO nanoparticle films on FTO-glass substrates, respectively. The microporous structure scatters visibly more light compared to the nanoporous morphology. (b) Transmission spectra of the two electrode types in (a) versus a FTO-glass reference. Integrating sphere measurements record the total transmitted light, while the UV-VIS spectrometer data (dashed line) only captures the light transmitted along the optical axis  omitting the scattered transmitted light. The microporous  electrode transmits less light because of increased scattering caused by the internal interfaces which leads to increased  backscattering.}
	\label{fig:subfigOptics}
\end{figure}

\footnotesize
{\sffamily
\textbf{Making of inverse opal ITO electrodes.}
The microporous electrodes were templated by polystyrene microshperes with an average diameter of 40\,\textmu m (Dynoseeds TS 40 from Microbeads). For each electrode, 450\,\textmu l of a 5\,\%wt suspension in deionised water were dried on a conductive FTO glass substrate (8 Ohm/sq, Sigma Aldrich). A homogeneous area (ca.\ 2\,cm$^2$) of closed-packed spheres was achieved when drying the suspension at 30\degr C on the substrate between two tightened aluminium sheets sandwiching a silicone O-ring, and with a hole in the upper sheet of the same size as the O-ring inner diameter. The frame can be used for convective assembly of colloids, but the large microspheres here settled on the substrate quickly, decreasing the influence of meniscal forces on the assembly. The bead structure was annealed for 10\,min at 130\degr C on a hotplate. Nanoparticles were derived from a commercial ITO nanoparticle dispersion ($<$100 nm) 30\,\%wt in isopropanol (IPA) (Sigma Aldrich) by drying and re-suspending in absolute ethanol (Sigma Aldrich) to result into a 10\,\%wt suspension. Ethanol was better suited as a solvent because of its wetting properties with the polystyrene opal. IPA leads to an increased coverage of the template top, leaving fewer entry points for the microorganisms. The polystyrene opal was placed on a hotplate at 45\degr C for quick drying and filled with 3 times with 25\,\textmu l of ITO suspension. The template was burned out on a hotplate inside a fume hood by heating the anodes to 500\degr C, which simultaneously sinters the nanoparticles, finishing the electrode. For this purpose, the following heat ramp procedure was used: 3h heat ramp from room temperature to 300\degr C holding the temperature for 1 minute;  10\,min ramp to 325\degr C holding for 5\,min; 10\,min ramp to 375\degr C for 5\,min; 10\,min to 450\degr C for 15\,min; 10\,min to 500\degr C for 15\,min; off.  This method can be found in detail on \href{http://docubricks.com/viewer.jsp?id=22282785798926336}{DocuBricks}.\\
For impedance spectroscopy testing, microporous electrode samples were prepared on ITO-coated glass substrates (to avoid an additional material interface) with the same protocol and annealed. They were then placed on upside down on a second ITO-glass substrate which was freshly blade coated with ITO nanoparticle paste (see below), and annealed again to sinter the paste and provide a good contact. After the annealing, electrodes were filled with room-temperature curing epoxy (EPO-TEK optical epoxy, 301-1LB kit) to provide handling stability. 
\ins{Four samples containing 130-150\,\textmu m thick microporous ITO layers made from the same nanoparticles had resistance values of $110-165\,\Omega$}.

\textbf{Making of flat ITO electrodes.}
Plain ITO on PET electrodes with a surface resistivity of 100\,Ohm/sq. were used as purchased from Sigma Aldrich. Nanoparticle film electrodes were made from the same commercial ITO nanoparticle dispersion and FTO glass substrate as the microporous electrodes. 2.5\,g equivalent of nanoparticles were mixed with 10.7 ml terpineol (a-Terpineol 96+\%, SAFC supply solutions). The IPA was evaporated off in a rotary evaporator at 55\degr C and vacuum pumping. FTO glass was taped with kapton tape (Polyimide Tape from RS components) as distance spacer, and the ITO-particle-terpineol paste was applied in between the tape strips. Excess paste was removed by manually pushing the side of a glass pipette rod over the spacers (blade-coating) to yield a plain film. The electrode was left to settle at room temperature for ca.\ 20 minutes. The kapton tape was removed manually. The electrodes were annealed with the same heat ramp protocol described for microporous samples.  This method can be found in detail on \href{http://docubricks.com/viewer.jsp?id=22282785798926336}{DocuBricks}. 
For impedance spectroscopy testing, ITO nanoparticle films were prepared on commercial ITO-coated glass substrates (to avoid an additional material interface) with the same blade coating protocol, and via spin-coating at different speeds. Commercial ITO-coated glass samples without any nanoparticle coating were annealed as the other samples, to provide an appropriate reference point despite the deterioration of ITO conductivity during the high temperature treatment. The mechanical junction of two individually annealed film samples provided a less reliable contact than the sintered microporous films, possibly causing a slight overestimation of their ohmic resistivity.

\textbf{Characterisation of porous electrodes.}
The transmission spectra in Fig.\,\ref{fig:subfigOptics} of the film were recorded with a Varian Cary 300 UV-Vis spectrophotometer, and with an Ocean Optics FOIS-1 integrating sphere  together with an Ocean Optics USB4000 spectrometer; both using uncoated FTO-glass as optical reference measurement (blank). The sheet resistivity of the nanoparticle film electrodes was determined on a non-conductive glass substrate with four point probe (S302 Lucas Labs and Keithley 2400) to be 100 $\pm$ 10 Ohm*cm$^{-2}$. The thickness of the films was found to be 8.9\,$\pm$ 0.6\,\textmu m by scanning over scratches with a Dektak 6M stylus profiler. To probe the conductivity laterally, and through the depth of the electrode film and compare these values to the microporous inverse-opal structure, sandwich samples were built either by clipping two nanoparticle film samples on ITO glass substrate on top of each other with two paper-clips, or by sintering a microporous electrode onto a film (details in respective electrode making method sections). The overlap area probing the material films was about 0.25\,cm$^2$. The clean end pieces of the ITO-glass substrates were contacted with silver paste and crocodile clips, and electrode impedance spectroscopy measurements were performed across the enclosed films, using the following parameters on a Biologic SP-300 Potentiostat: Scan at 0\,V from 1\,MHz to 1\,Hz, with 40 points per decade, a sinus amplitude of 5\,mV, waiting 0.1\,periods before each frequency. No imaginary parts of the Nyquist plot were found that could indicate a non-ohmic resistance behaviour of the nanoporous or microporous films. The presented resistance data in this study is thus simply an average of the (noisy but stable) real part of the measurements.

\textbf{Bacteria growth and quantification.}
\emph{Synechocystis} sp.\ PCC 6803 and \emph{Nostoc punctiforme} cells were routinely cultured in BG11 medium supplemented with about 1\,mM NaHCO$_3$ and maintained in sterile conditions at 30\degr C under continuous moderate light of 40-50\,\textmu mol photons m$^{-2}$ s$^{-1}$ and shaking at 160\,rpm. Before applying cells to electrochemical devices, they were concentrated via centrifugation (ca.\ 2000\,g for 10\,min). A small volume of cells (120 - 160\,\textmu l) containing 134\,nmol chlorophyll were pipetted into the devices pre-filled with BG11 medium and after un-inoculated reference measurements. 
The heavy cell suspension quickly settled onto the anode to form a biofilm before the cell-suspension can mix with the bulk electrolyte.
The chlorophyll concentration was measured by extracting it from the cell suspension in 99.8\,\% methanol   (Sigma-Aldrich) and then \rem{subtracting the 750\,nm O.D. value from the 680\,nm O.D. value and multiplying by 10.854} \ins{calculating the chlorophyll-a concentration from two optical densities, as described previously}~\autocite{Porra1989}.

\textbf{Biophotovoltaic devices.}
The devices were printed on a Projet 3500HD Max 3D printer with Visijet M3X acrylic material and an accuracy of 0.025\,mm. The support material (VisiJet S300) was washed out with hot sun-flower oil and then IPA. Upright air-cathode windows (carbon paper with Pt-nanoparticles, Alfa Aesar 45372 Hydrogen Electrode/Reformate) were sealed into the 3D printed device with dental silicone (Zhermack, Elite HD+ Super Light Body). The anodes were tightened in the device manually with screws and sealed with a nitrile O-ring. A good contact between the anode edges and the connector-clamp was ensured by conductive silver paste.

\textbf{BPV operation and measurements.}
The BPV devices were loaded with the different anode types and filled with a 10\,ml volume of BG11 medium (containing small amounts of phosphate) and phosphate buffer (DPBS 10x, D1408 Sigma-Aldrich, pH~7.2) in a mixing ration of 9:1.07 to obtain an overall phosphate buffer concentration of 10\,mM. Experiments referred to as `no added phosphate buffer' used devices filled only with BG11 medium.  The electrochemical measurements were performed with a potentiostat (MultiEmStat by PalmSens) with 12 independent channels. The potentiostat applies a voltage and measures the resulting current with a high resolution of down to 1\,nA. Reference measurements were conducted with the connected devices after at least 6 hour waiting time (when chronoamperometry measurements seemed to have reached equilibrium) and before adding cyanobacteria. Then, 134\,nmol chlorophyll-a equivalent of cyanobacteria were added to each device (see Bacteria growth and quantification).  Measurements were conducted again after a settling time of at least 8\,h and usually for a duration of one to three days. The illumination periods with high light intensities were limited to five or seven minutes, while more moderate photon flux densities of ca.\ 500\,\textmu mol/m$^2$/s were used for longer exposures. In the case of the custom LED white light source (Fig.\,\ref{fig:subfigIllumination}), the moderate illumination compares to the world average sun intensity reaching ground level within the visible light spectrum. The temperature of the devices was held stable at 25\degr C by water cooling or heating of the hollow devices with an external water bath.\\
The number of measurement channels per anode type used to obtain the presented averages and standard deviations vary slightly because the devices that shorted due to water condensation or leakage were not considered. During BPV device measurements with \emph{Synechocystis}, all four devices each with micro- and nanoporous electrodes and three devices with non-porous electrodes delivered uninterrupted data.  In half-cell mode, averages were formed of three microporous electrode bearing devices, four with NP-films and three with non-porous electrodes. For \emph{Nostoc} measurements, the numbers were two, four, and two respectively. No increase in \emph{Nostoc} electro-activity was observed after several days in the device, in contrast to the literature~\autocite{Sure2016}.\\
\rem{Shewanella fuel-cell measurements.}
Measurements on \emph{Shewanella oneidensis} bacteria were conducted in half-cell mode at an anode potential of 0.2\,V vs.\ an  Ag/AgCl reference electrode in fresh LB medium without additional phosphate buffer.  The same amount of concentrated cell suspension  was added to each device (1400\,\textmu l each, cells were separated by repeated pipetting qith a 1\,ml tip), and transmission spectra were recorded at different cell densities as a point of reference for cell numbers, see Fig.\,\ref{fig:subfigODShewanella}. Data was generated from four devices with micro- and nanoporous  electrodes each, and two non-porous ITO electrodes. The devices were covered with Parafilm before measuring, which reduced the oxygen supply. The Parafilm did not create fully anaerobic conditions, since during cyclic voltammetry measurements, the presence of residual oxygen influx could still be detected.

\textbf{Microscopy.}
Optical microscopy was conducted with an Olympus BX60 microscope and Olympus UMPlanFl 20x and 100x objectives. Images at manually adjusted different focal depths were taken with an AxioCam MRc 5 (Zeiss) camera and Z-stacked with CombineZP software. Slight adjustments were made to the colour balance and contrast of the image.
For electron microscopy images, a Leo Gemini 1530 VP SEM was used with a Schottky-emitter consisting of a zirconium oxide coated tungsten cathode and an in-lens secondary electron detector.

\textbf{Data availability.}
The authors declare that the data supporting the findings of this study will be made available in a public repository to which a permanent reference will be included in the paper in time for publication.

}
\begingroup
\setlength\bibitemsep{0pt}
\printbibliography[heading=subbibliography,title=\textsf{References}]
\endgroup
\end{refsection}


\section*{\textsf{Acknowledgements}}
\footnotesize
{\sffamily
We thank Antonio Abate for advice on conductivity testing across the films as well as interpreting the impedance spectroscopy results and Erika Eiser for discussion and the provision of colloids. TW is grateful for funding from the Winton Fund for physics of sustainability, Mott Fund, Cambridge Trust (CHESS), and the EPSRC CDT in Nanoscience and Nanotechnology. DH thanks Evangelisches Studienwerk -- Villigst e.V.\ for funding during this work. PB and CJH thank the Leverhulme Trust for financial support. US acknowledges funding from the Adolphe Merkle Foundation.

\section*{\textsf{Author contributions}}
TW, US, PB, and CH conceived the project and designed the experiments. TW collected and analysed all included data and performed imaging. TW \& DH designed and built electrochemical devices, light source, and method for microporous electrodes with advice from PB \& US. TW wrote the manuscript, further improved by all authors.

\section*{\textsf{Additional information}}
Competing interests:
The authors declare no competing financial interests.
}


\onecolumn
\section*{Supporting Information for}
\beginsupplement
{\bf \LARGE Porous translucent electrodes enhance current generation from photosynthetic biofilms}\\
\vspace{0.05cm}\\
{\Large Tobias Wenzel, Daniel Haertter, Paolo Bombelli, Christopher Howe, Ullrich Steiner}
\noindent
\vspace{0.5cm}\\
\normalsize

\begin{figure}[H]
	\centering
	\includegraphics[width=0.65\textwidth]{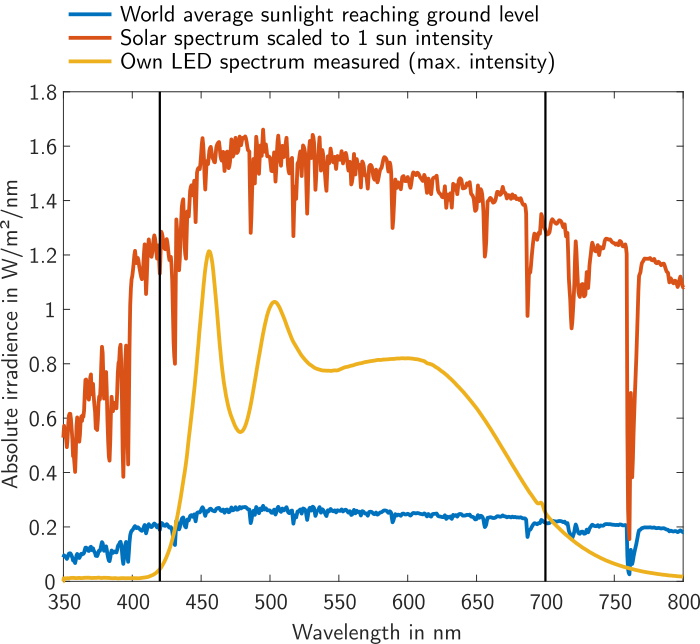}
	\caption{Illuminating light source. Spectrum of custom LED light source (yellow) at maximum intensity compared to the solar spectrum (red) (reference AM 1.5 Spectrum by ASTM, global tilt) at intensities of 1000\,W/m$^2$ (solar simulator standard) and the world average sun intensity reaching ground level (blue).}
	\label{fig:subfigIllumination}
\end{figure}

\begin{figure}[H]
	\centering
	\includegraphics[width=0.85\textwidth]{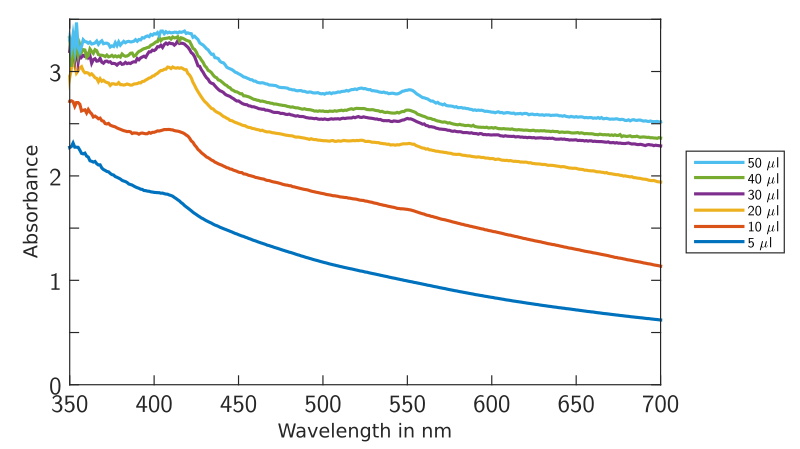}
	\caption{UV-VIS absorption spectra of liquid \textit{Shewanella} cell-culture at different dilution ratios. 
	The amount of concentrated cells between 5\,\textmu l and 50\,\textmu l, as indicated in  legend, was added to 1\,ml of LB medium (reference solution) for each measurement. 140\,\textmu l of concentrated cell-culture was used per device for electrochemical measurements. This data serves as a reference for the amount of cells added, in absence of an equivalent protocol for the determination of chlorophyll concentration.}
	\label{fig:subfigODShewanella}
\end{figure}

\end{document}